\def\siri{{\sl SIRIUS}}

\documentclass{iau}
\usepackage{graphicx}

\title[Massive Stars in Galactic Center] 
{Unveiling the Massive Stars in the Galactic Center}

\author[Dong, H. et al.]   
{H. Dong$^1$, 
 J. Mauerhan$^{2,3}$, M. R. Morris$^4$, Q. D. Wang$^5$, A. Cotera$^6$}

\affiliation{$^1$ National Optical Astronomy Observatory, Tucson, AZ
  85719 \\ $^2$ Department of Physics, University of California,
  Berkeley, CA 94720 \\ $^3$ Steward Observatory, University of
  Arizona, Tucson, AZ 85719 \\ $^4$
Department of Physics and Astronomy, University of California, Los
Angeles, CA 90095 \\ $^5$ Department of Astronomy, University of Massachusetts,
Amherst, MA 01003 \\ $^6$ SETI Institute, Mountain View, CA 94043 \\ email: {\tt hdong@noao.edu}}

\pubyear{2013}
\volume{303}  
\pagerange{???--???}
\jname{The Galactic Center: Feeding and Feedback in a Normal Galactic Nucleus}
\editors{A.C. Editor, B.D. Editor \& C.E. Editor, eds.}
\begin{document}

\maketitle

\begin{abstract}
We present our recent efforts to unveil and understand the origin of massive stars outside 
the three massive star clusters in the Galactic Center. From our
HST/NICMOS survey of the
Galactic Center, we have identified 180
Paschen-$\alpha$ emitting sources, most of which 
should be evolved massive stars with strong optically thin stellar
winds. Recently, we obtained Gemini GNIRS/NIFS H- and K-band spectra of eight massive
stars near the Arches cluster. From their radial velocities, ages and
masses, we suggest that in our sample, two stars are previous
members of the Arches cluster, while other two stars embedded in the
H1/H2 H{\small \rm II} regions formed in-situ. 
\end{abstract}

\firstsection 
\section{Motivation}
The Galactic Center (GC) is the only galactic nucleus, in which we can study the
co-evolution of nuclear star formation activity, a nuclear star cluster 
and a super-massive black hole (SMBH) in great detail. During the last
two decades, a SMBH, Sgr A* (\cite{ghe08,gil09}), and a nuclear cluster, 
the Center cluster (\cite{gen03}), are
indeed found in the GC. Meanwhile, the central 300 pc of the GC,
known as, the Central Molecular Zone (CMZ), includes $\sim$4$\times 10^7~M_{\odot}$ molecular
clouds (\cite{mor96,mol11}), which are forming stars
with a rate of about 0.03$M_{\odot}$/yr (\cite{lon13}). Because of the
proximity, we can detect individual massive stars and study their
interaction with the interstellar medium (ISM).

One important question that we want to address 
is how many stars formed during the last 10 Myr in the
CMZ and whether the energy released by these stars could explain 
the Fermi Bubbles 
(\cite{su10}, \cite{cro11}). 
Only three young, massive and compact star clusters, the Arches,
Quintuplet and Center clusters (2-7 Myr old, \cite{fig99,fig02,gen03}), have been
identified in ground-based near-infrared (NIR) surveys of the
Galactic Center. Young massive stars outside of these three clusters have largely been unknown. The slow progress in detecting these
`field' stars is largely owed to the lack of an efficient way to distinguish 
young massive stars and evolved low-mass stars through their NIR
colors. 

Our HST/NICMOS Paschen-$\alpha$ GC
survey (\cite{wan10}) provides a systematic way to identify evolved
massive stars (EMS) having strong stellar winds. We
mapped out the central 39\arcsec$\times$15\arcsec\ of the GC with NIC3 in two
narrow-band filters: F187N (on-line, 1.87$\mu$m) and F190N (adjacent
continuum, 1.90$\mu$m) (\cite{don11}). Combined with archived HST/NICMOS 
NIC2 observations of the three star clusters at the same bands, we
identify 180 Paschen-$\alpha$ emitting sources (PESs) in our survey
area (see Fig. 1a), which have strong line emission in the F187N band and could be 
EMSs (\cite{don12}). 
Half of these stars lie outside the three clusters and are
scattered throughout the GC. Unlike those in the clusters, most
of these `field' stars still don't have available spectroscopic
identifications. 


\section{Near-Infrared Photometry of Paschen-$\alpha$ Emitting
  Sources}
Before obtaining spectroscopic observations of `field'
PESs, we use the existing NIR photometry to study their properties. We
employ the SIRIUS (\cite{nag03}) and 2MASS (\cite{skr06}) source
catalogs. 
Fig. 1b and
Fig. 1c show the color-color and magnitude-color diagrams for our 164
Paschen-$\alpha$ emitting sources with available NIR photometry. Most
of our PESs are very red and
indeed within the GC. We also estimate their extinction using the NIR
colors (\cite{don12}). 

\begin{figure}[b]
\begin{center}
 \hspace*{-1cm}\includegraphics[width=5.0in]{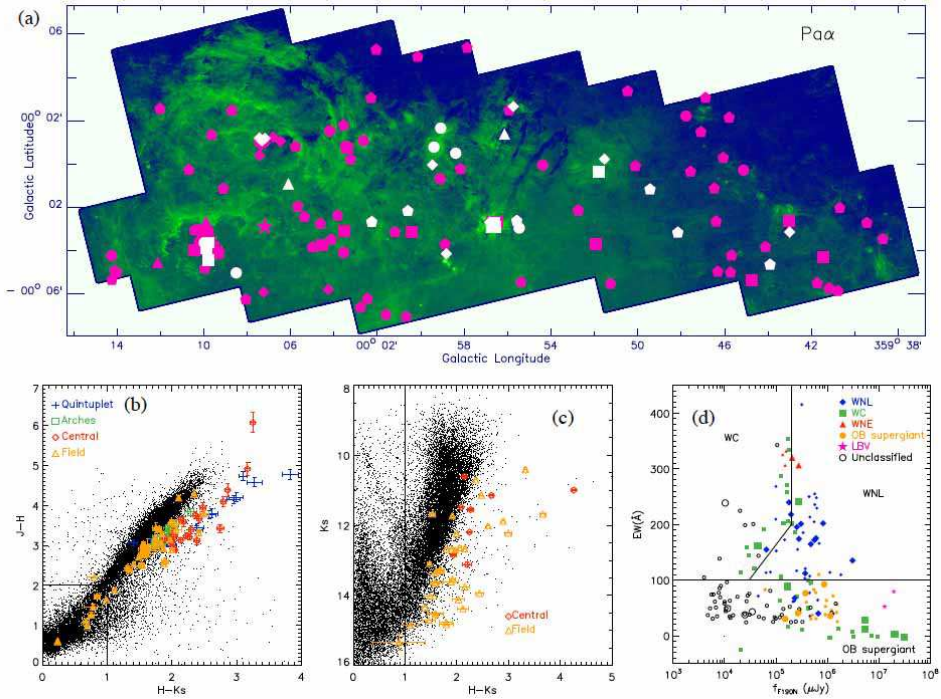} 
 \caption{ (a) 180 PESs overlaid on the mosaic image of Paschen-$\alpha$
   intensity (`diamond': WNL stars, `square': WC stars, `triangle':
   WNE stars, `circle': OB supergiants, `star symbols': LBV stars). The `pentagon' symbols represent those Paschen-alpha without available spectroscopic identifications. The filled symbols indicate ones with X-ray counterparts. (b) and (c) are infrared
   color-color and magnitude-color diagram of
   119 and 45 PESs, compared to the field stars 
from the \siri\ survey
 (black small dots). Thick black lines delineate the region occupied by foreground
 stars (H-K $<$1 and J-H $<$ 2). (d) The combined Paschen-$\alpha$ EW vs. $f^o_{F190N}$ plot 
of stars of various types. Lines are drawn to roughly show the regions dominated by 
WNL, WC and OB supergiants. The stars with larger symbols are the ones with X-ray counterparts.}
   \label{fig1}
\end{center}
\end{figure}

Fig. 1d shows the equivalent width (EW) in the F187N band versus the
intrinsic F190N intensity of our PESs. We empirically divide this figure into three
parts according to their spectral types. The EWs of the WN stars are
large and seem to be correlated with the intrinsic F190N
intensity. On the other hand, the EWs of the WC stars
span a large range, which could be due to surrounding 
hot dust shells enhancing the
F187N continuum and thus reducing the EWs. Compared to the WN stars, 
the EWs of OB supergiants are smaller, indicating weaker stellar
winds. Therefore, those PESs
without available spectroscopic identification most likely represent a combination
of WC stars and OB supergiants. 

\section{The Origin of Paschen-$\alpha$ Emitting
  Sources}
These `field' PESs could: 1) be the leftover of dissolved
star clusters; 2) be ejected out of the three existing massive star
clusters through three-body interactions; and/or 3) form {\sl
in-situ}. 

In order to understand their origin, 
we measured the radial velocities of eight OB supergiants and O If$^+$
stars
 within 10\arcmin\ of the Arches cluster with Gemini GNIRS/NIFS NIR
 spectra. 
With the Gemini/GNIRS long-slit spectra, we also detects the
Br$\gamma$ line emission from nearby
H{\small \rm II} regions. We further complement our data with previous radio
H92$\alpha$ (\cite{lan02}) and CS observations (\cite{tsu99}) for the radial
velocity of nearby ionized gas and molecular clouds. By comparing
their unreddened NIR colors with Geneva stellar evolutionary tracks,
we conclude that these
eight stars are younger than 5 Myr and have masses $>$25
M$_{\odot}$. Therefore, they could not
be from dissovled clusters, which should have lifetimes greater than 7
Myr (\cite{por02}).  

We mark the locations of the eight stars in Fig. 2a.
 Combined with the proper motion derived from their
projected distances, the 3-D velocities of the stars relative to the
Arches cluster (assuming that it is their origin) are 60-140 km/s, still within the range
predicted by three-body interactions of massive stars (\cite{gva11}). On
the other hand, most of these stars are associated with in local
extended Paschen-$\alpha$ emission and could form {\sl in-situ}. 

We compare the radial velocities of these stars with those of
nearby ionized gas, molecular clouds and the Arches cluster in
Fig. 2b. All the stars are blueshifted, relative to the Arches
cluster, by $>$ 50 km/s. The different radial velocities of P97/P107
and nearby molecular clouds indicate that they are just runaway
stars, while near P98/P100, the morphologies of ionized gas don't support
that they are young star formation regions. Because 
P97 and P98 have spectra and ages similar to those of O
If$^+$ stars in the Arches cluster, they could be previous
members of the Arches cluster. Because P100 and P107 are
relatively older than the Arches cluster (Dong et al, in preparation), future proper motion are
needed to confirm their relationship. Combined with
three WNh stars within 1-2 pc of the Arches cluster suggested by
\cite[Mauerhan et al. (2010)]{mau10}, there are five `field' EMSs which could have been ejected from
the Arches cluster. 

\begin{figure}[b]
\begin{center}
 \vspace*{-1.0 cm}\hspace*{-1cm}\includegraphics[width=6.5in]{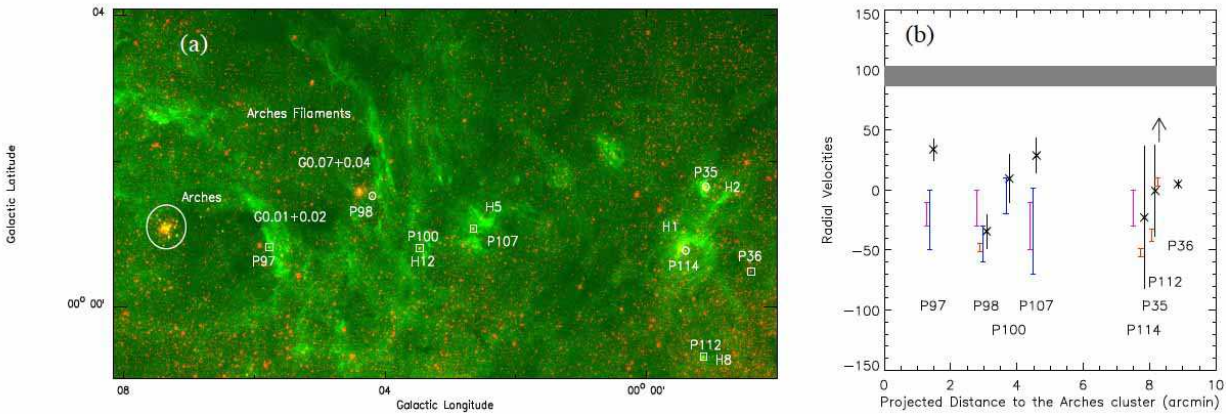} 
 \caption{Left: eight evolved massive stars overlaid on
   the color image; red: HST F190N image and green: 
Paschen-$\alpha$ emission. We mark the
 location of the Arches cluster and different H{\small \rm II}
 regions. Right: the 
radial velocities of our massive stars (`crosses'), nearby ionized
   gas (the blue and red lines on the left side of individual
   `crosses') and molecular clouds (the cyan lines). 
The x-axis is the projected distance from the
   massive stars to the Arches cluster. The grey region is the 
radial velocity of the Arches cluster.}
   \label{fig2}
\end{center}
\end{figure}

A complete census of these deserters is
essential to understanding the properties of the Arches
cluster. In addition to producing these runaway stars, three-body interactions 
may also produce tightly bound massive binaries. Several of these bainries probably 
don't have enough velocities to have escaped from the Arches
cluster. Indeed, three massive binaries have been found in
the Arches cluster (\cite{wan06}), through their X-ray emission from
the wind collision
zones. Of course,  there may be more massive binaries in the Arches cluster,
which produce too weak X-ray emission to have been detected. If such
binaries are considered as single stars, their ages/masses could be
underestimated/overestimated. Therefore, detecting the ejected massive
stars can provide useful constraints on the dynamic processes of the cluster.

The remainnig four stars could be unrelated to
the Arches cluster. No stars in the Arches cluster have 
spectral types similar to P112, and the O If$^{+}$ stars in the Arches cluster
are more massive than P36. P114 and P36 are embedded in the H1 and H2 H{\small
  \rm II} regions, which have shell-like structures. P36/P114 
and the nearby -30-0 km/s molecular cloud
 have similar radial velocities, redshifted in respect to  
surrounding ionized gas. The lack of no H$_2$ 2.121
 $\mu$m line indicates that the shell-like structures
 can't be explained by a bow shock model, but perhaps by a pressure-driven
 model. Therefore, P36/P112 are not runaway stars and 
could have formed {\sl in-situ}. According to the toy model of the
 molecular clouds in the CMZ proposed by \cite{mol11}, about 2 Myr
 ago, the nearby -30-0 km/s molecular cloud could have passed its
 closest approach to Sgr A* and
 the star formation was triggered by the strong stellar winds
 from the Central cluster, and also the possible feedback from Sgr
 A*.



\begin{thebibliography}{}
\bibitem[Ghez et al. 2008]{ghe08} Ghez, A.~M., Salim, S., 
Weinberg, N.~N., et al.\ 2008, ApJ, 689, 1044 

\bibitem[Gillessen et al. 2009]{gil09} Gillessen, S., 
Eisenhauer, F., Trippe, S., et al.\ 2009, ApJ, 692, 1075 

\bibitem[Genzel et al. 2003]{gen03} Genzel, R., Sch{\"o}del, 
R., Ott, T., et al.\ 2003, ApJ, 594, 812 

\bibitem[Morris \& Serabyn 1996]{mor96} Morris, M., \& Serabyn, E.\
  1996, ARA\&A, 34, 645 

\bibitem[Molinari et al. 2011]{mol11} Molinari, S., Bally, 
J., Noriega-Crespo, A., et al.\ 2011, ApJL, 735, L33 

\bibitem[Longmore et al.(2013)]{lon13} Longmore, S.~N., 
Bally, J., Testi, L., et al.\ 2013, MNRAS, 429, 987 

\bibitem[Su et al. 2010]{su10} Su, M., Slatyer, T.~R., 
\& Finkbeiner, D.~P.\ 2010, ApJ, 724, 1044 

\bibitem[Crocker 
\& Aharonian(2011)]{cro11} Crocker, R.~M., \& Aharonian, F.\ 2011, Physical Review Letters, 106, 101102 

\bibitem[Figer et al. 1999]{fig99} Figer, D.~F., McLean, 
I.~S., \& Morris, M.\ 1999, ApJ, 514, 202

\bibitem[Figer et al. 2002]{fig02} Figer, D.~F., Najarro, 
F., Gilmore, D., et al.\ 2002, ApJ, 581, 258 

\bibitem[Wang et al. 2010]{wan10} Wang, Q.~D., Dong, H., 
Cotera, A., et al.\ 2010, MNRAS, 402, 895 

\bibitem[Dong et al. 2011]{don11} Dong, H., Wang, Q.~D., 
Cotera, A., et al.\ 2011, MNRAS, 417, 114 

\bibitem[Dong et al. 2012]{don12} Dong, H., Wang, Q.~D., 
\& Morris, M.~R.\ 2012, MNRAS, 425, 884 

\bibitem[Nagayama et al. 2003]{nag03}Nagayama, T., Nagashima, C.,
  Nakajima, Y., Nagata, T., et al., SPIE, 4841, 459N

\bibitem[Skrutskie et al. 2006]{skr06}Skrutskie, M. F., Cutri, R. M.,
  Stiening, R., Weinberg, M. D., et al., 2006, AJ,
  131, 1163S


\bibitem[Lang et al. 2002]{lan02} Lang, C.~C., Goss, W.~M., 
\& Morris, M.\ 2002, AJ, 124, 2677

\bibitem[Tsuboi et al. 1999]{tsu99} Tsuboi, M., Handa, T., 
\& Ukita, N.\ 1999, ApJS, 120, 1  

\bibitem[Gvaramadze 
\& Gualandris 2011]{gva11} Gvaramadze, V.~V., \& Gualandris, A.\ 2011,
MNRAS, 410, 304 

\bibitem[Portegies Zwart et al. 2002]{por02} Portegies 
Zwart, S.~F., Makino, J., McMillan, S.~L.~W., 
\& Hut, P.\ 2002, ApJ, 565, 265 

\bibitem[Mauerhan et al. 2010]{mau10} Mauerhan, J.~C., 
Cotera, A., Dong, H., et al., 2010, ApJ, 725, 188 

\bibitem[Wang et al. 2006]{wan06} Wang, Q.~D., Dong, H., 
\& Lang, C.\ 2006, MNRAS, 371, 38 
\end{thebibliography}
\end{document}